\documentclass[conference,final]{IEEEtran}
\usepackage{blindtext, graphicx}
\usepackage{url}

\begin{document}

\title{Will humans even write code in 2040 and what would that mean for extreme heterogeneity in computing?}

\author{\IEEEauthorblockN{Jay Jay Billings, Alexander J. McCaskey, Geoffroy Vallee, and Greg Watson}\\
\IEEEauthorblockA{Oak Ridge National Laboratory\\
PO Box 2008 MS6173\\
Oak Ridge TN 37830\\
Email: billingsjj@ornl.gov\\
Twitter: @jayjaybillings}
}

\maketitle

\begin{abstract}

Programming trends suggest that software development will undergo a radical change in the future: the combination of machine learning, artificial intelligence, natural language processing, and code generation technologies will improve in such a way that machines, instead of humans, will write most of their own code by 2040. This poses a number of interesting challenges for scientific research, especially as the hardware on which this machine-generated code will run becomes extremely heterogeneous. Indeed, extreme heterogeneity might drive the creation of this technology because it will allow humans to cope with the difficulty of programming different devices efficiently and easily.

\end{abstract}

\underline{Notice of Copyright:} This manuscript has been authored by UT-
Battelle, LLC under Contract No. DEAC05-00OR22725 with the U.S. Department of
Energy. The United States Government retains and the publisher, by accepting
the article for publication, acknowledges that the United States Government
retains a nonexclusive, paid-up, irrevocable, world-wide license to publish or
reproduce the published form of this manuscript, or allow others to do so, for
United States Government purposes. The Department of Energy will provide
public access to these results of federally sponsored research in accordance
with the DOE Public Access Plan (http://energy.gov/downloads/doe-public-access-plan).

\section{Introduction}
Consider the following simple question that might be posed as a computational problem between two researchers: \textit{Given my morning cup of Starbucks coffee, under standard assumptions, what is the temperature of the coffee after ten minutes?}

The second researcher, who is charged with writing a code to calculate the temperature, needs to determine the meaning and values of a number of properties (the ``standard assumptions'') but will otherwise quickly recognize that the answer to this problem is the solution to the time-dependent heat equation or something that given enough data could be computed by interpolation. The second researcher will follow one or the other of two time-honored procedures---calculating the theoretical value or computing the value based on measured data---and report the result to the first researcher along with some implied or explicit amount of error. This is not a new scenario, but given new developments in machine learning, artificial intelligence, natural language processing, and code generation, \textit{does the second researcher necessarily have to be human?}

Present programming trends and research directions suggest that by 2040 the answer to this question might very well be no and that machine-generated code (MGC) might be as common as artificial intelligence in devices today or as self-driving cars in the next couple of years. The major technologies that will drive the creation and adoption of MGC already exist, either at research institutions or in the marketplace. A number of efforts exist that are attempting to streamline how these technologies can be used more efficiently either to write new code from scratch or to enable learning at a faster rate. The Defense Advanced Project Agency's (DARPA's) Probabilistic Programming for Advancing Machine Learning (PPAML) program is developing new technologies that improve machine learning for questions such as the previous example on temperature \cite{noauthor_probabilistic_nodate}. Both DeepCoder \cite{balog_deepcoder:_2016} and AutoML \cite{automl} use machine learning to produce executable code. Ontology generation tools, including DOG4DAG, can semiautomatically generate entire knowledge bases that humans might not be able to generate on their own because of time constraints \cite{wachter_dog4dag:_2012}. Code-generation technologies from the Eclipse Foundation, including the Eclipse Modeling Framework and Sirius, can not only generate the entire data hierarchy for a project but also the entire user interface and middle layer \cite{steinberg_emf:_2008,sirius}. If humans do need to write some of the code, they might find that they spend more time using autocomplete and code recommendation features than on writing new lines on their own \cite{codeRecommenders}. The final piece of this puzzle is that application programming interfaces in scientific libraries are becoming standardized and require only some understanding of the problem domain, not the library itself (c.f., the examples for the Modified Finite Element Method [MFEM] project \cite{mfem}).

The pressing question then is not will MGC become common place but what should the scientific community do when it does. This work examines that question in the context of extreme heterogeneity, which might simultaneously drive the need for, support, and benefit from MGC.

\section{Key Challenges}
In the context of scientific research, the challenges around MGC are more technological than sociological. From a sociological perspective, it is tempting to say that MGC will face the same fate as both domain-specific languages (DSLs) and high-productivity computing system (HPCS) languages, such as Chapel, which largely failed to change scientific computing. However, using DSLs and HPCS languages comes at the cost of learning those languages and installing the associated tools. MGC does not, and adoption will be quick.

Arguably the most pressing technological challenge for extreme heterogeneity is programming all the different hardware types efficiently. For a human user, this requires a high-level language or, if MGC is employed, an appropriate natural language processing interface. If a machine writes the code are there more efficient languages or abstractions for machine-to-machine communications, especially across hardware types? Some early results from Facebook this year suggest that machines are capable of developing their own more efficient methods of communication, even with some of the restrictions (letters) of natural language imposed \cite{wilson_ai_2017}. MGC is already used in hybrid classical plus quantum computing systems with the eXtreme-scale ACCelerator (XACC) project \cite{mccaskey_extreme-scale_2017}.

Another important challenge is how to allocate hardware resources for the purposes of authoring code. One great benefit of heterogeneity is optimizing resources allocation for specific tasks by exploiting natural fits with hardware types and problems. Assuming that MGC is the future of programming, are some types of hardware a better choice for authoring code than others? For example, neuromorphic processors such as TrueNorth \cite{merolla_million_2015} are better at recognizing patterns than performing double precision arithmetic, which could be used for choosing algorithms and implementation details, and quantum computers are better at optimization, which could be adapted for code optimization.

It is also important to consider the possibility that ``native'' machine languages and other natural consequences of MGC allow machines to find optimal solution strategies or fix areas of concern far better than humans. This has been demonstrated at a small scale with so-called ``adaptive'' applications that use machine learning to optimize solver parameters and finite element meshes.

\section{Research Directions}
Several research directions were alluded to previously. For example, it will be important to produce compilers and other tools that can cross-compile on heterogeneous systems and do so efficiently in ways that are ``native'' for machines and to do so under optimization conditions where compilation could be happening thousands or millions of times per second on specialized hardware for the artificial intelligence to test and learn. It will also be important to look at heterogeneous hardware and ask questions about what types of hardware should be allocated for MGC in various stages such as problem recognition and natural language processing, code definition, code generation, and code optimization. There are a number of other interesting areas of research as well. Would it be possible for artificial intelligence to automatically learn how to use a new hardware component without instruction from a human? If so, would the best method be brute force or are there better discovery techniques? The importance of native machine languages is also an interesting topic of research because, in spite of unnecessary fears that humans will not be able to understand the languages, significant efficiency might be gained by taking humans, scripts, files, and pipes out of the programming and compilation loop. This is a natural complement to machine agents running across hardware types that need to quickly learn how to communicate and share work.

MGC and extreme heterogeneity combined have strong implications on the future of reproducibility and repeatability. As hardware changes, answers change. The ultimate goal is to narrow the gap between repeatability and reproducibility so that new results are as good as if not better than the original results, even if the values are not exactly the same. MGC seems to fit this neatly because it makes it possible to describe the higher level concepts accurately and easily while efficiently using all the available hardware without a requirement that human users know the details or do the translation. This is an aspect of MGC and extreme heterogeneity that applies to both individual codes and complex workflows since, presumably, many of the same techniques would be shared between the two. Efficient ways of leveraging MGC for repeatability and reproducibility across hardware types will be very challenging to investigate.

There are also a number of important research directions for MGC with respect to data on heterogeneous systems. While data often has the added benefit of being self-describing, the generation rate and diversity of data are both increasing exponentially. Significant research challenges exist for intelligent agents that can dynamically adapt to changing data types, allocate resources as needed, and be data centric instead of machine specific so that computation can move across systems without regard for back-end hardware (and even undergo analysis in transit). Self-replicating programs that could replicate copies of themselves to move to data instead of moving the data might actually be the most logical choice for the largest data sets. Research for this case could look at the application of artificial intelligence swarms that spawn millions of agents versus single large, MPI-like programs.

\section{Summary}
Extreme heterogeneity, along with the rest of the computing world, will be required to move with the demands of usability and productivity in interesting ways. This could be as simple as code being automatically written and compiled by natural language processing or as complicated as machines developing native languages and sharing work without human intervention. Machines writing code under human direction will only further improve our ability to explore the universe, enjoy life, and stream Netflix, especially if it saves us the trouble of learning how to make extremely heterogeneous systems work together.

\section{Acknowledgments}

The authors are grateful for the feedback of David Bernholdt and Barney Maccabe, also of Oak Ridge National Laboratory (ORNL), for their generous input on the sociological aspects of HPCS languages and program/system synthesis, respectively.

This work has been supported by the US Department of Energy, the ORNL Director's Research and Development Fund, and by the ORNL
Undergraduate Research Participation Program, which is sponsored by ORNL and
administered jointly by ORNL and the Oak Ridge Institute for Science and
Education (ORISE). ORNL is managed by UT-Battelle, LLC, for the US Department
of Energy under contract no. DE-AC05-00OR22725. ORISE is managed by Oak Ridge
Associated Universities for the US Department of Energy under contract no.
DE-AC05-00OR22750.

\bibliographystyle{plain}
\bibliography{bib}

\end{document}